\documentclass[12pt]{article}
\usepackage{a4wide}
\usepackage{amssymb}
\usepackage{graphicx}
\begin{document}
{\renewcommand{\thefootnote}{\fnsymbol{footnote}}
\hfill  AEI--2005--091\\ 
\medskip
\hfill gr--qc/0504100\\
\medskip
\begin{center}
{\LARGE  Asymptotic Properties of Difference Equations\\
 for Isotropic Loop Quantum Cosmology}\\
\vspace{1.5em}
Martin Bojowald\footnote{e-mail address: {\tt mabo@aei.mpg.de}}
and Adam Rej\footnote{e-mail address: {\tt adamrej@box43.pl}}
\\
\vspace{0.5em}
Max-Planck-Institut f\"ur Gravitationsphysik, Albert-Einstein-Institut,\\
Am M\"uhlenberg 1, D-14476 Golm, Germany\\
\vspace{0.5em}
Division of Field Theory and Elementary Particle Physics, University of Silesia, Bankowa 14,40-007 Katowice, Poland\\
\vspace{1.5em}
\end{center}
}

\setcounter{footnote}{0}

\newtheorem{theo}{Theorem}
\newtheorem{lemma}{Lemma}
\newtheorem{defi}{Definition}
\newtheorem{prop}{Proposition}

\newcommand{\proofend}{\raisebox{1.3mm}{\fbox{\begin{minipage}[b][0cm][b]{0cm}
\end{minipage}}}}
\newenvironment{proof}{\noindent{\it Proof:} }{\mbox{}\hfill \proofend\\\mbox{}}
\newenvironment{ex}{\noindent{\it Example:} }{\medskip}
\newenvironment{rem}{\noindent{\it Remark:} }{\medskip}

\newcommand{\sgn}{{\rm sgn}}

%

\begin{abstract}
 In loop quantum cosmology, a difference equation for the wave
 function describes the evolution of a universe model. This is
 different from the differential equations that arise in
 Wheeler--DeWitt quantizations, and some aspects of general properties
 of solutions can appear differently. Properties of particular
 interest are boundedness and the presence of small-scale
 oscillations. Continued fraction techniques are used to show in
 different matter models the presence of special initial conditions
 leading to bounded solutions, and an explicit expression for these
 initial values is derived.
\end{abstract}

\section{Introduction}

Quantum gravity can in general be expected to provide a foundation
which is radically different from classical smooth background
geometries in that space and time should be discrete. Evolution
equations are then no longer given by differential equations in space
and time, but by difference equations which in general can be
approximated by differential equations only on large scales. This is
realized explicitly in loop quantum cosmology \cite{LoopCosRev}, based
on loop quantum gravity \cite{ALRev,Rov,ThomasRev}, where one obtains
difference equations for the evolution of a universe
\cite{cosmoIV}. This has physical implications, in particular
concerning the singularity issue, but also conceptual ones since
solutions of difference equations can generically behave differently
from those of differential equations.

In this paper we focus on properties of solutions of isotropic loop
quantum cosmology in the large volume region, where one expects
differential equations to provide a good approximation
\cite{SemiClass,FundamentalDisc}. There are several cases where the
boundedness of solutions is of interest, for instance when a
probability interpretation is used \cite{ClosedExp} or the physical
inner product is studied \cite{IsoSpinFoam}. From the continuum limit
one can draw generic expectations, as in quantum mechanics, since one
only needs to know whether regions of interest are classically
forbidden. In such a case, one expects exponential behavior of
solutions of a differential equation and bounded solutions only in
exceptional cases. However, since these issues can depend sensitively
on precise values of a solution, a direct analysis of the difference
equation without detour through the continuum limit is preferable.

This analysis is performed here with a general confirmation of the
expectations. As a spin-off of the investigation we obtain a
semi-analytical procedure which allows to determine initial values for
bounded solutions. The resulting formula is much stronger than a
numerical trial and error procedure, and does not only apply to those
few models where exact solutions are known.

\section{Isotropic loop quantum cosmology}
\label{lqc}

Dynamical properties of a universe in isotropic loop quantum cosmology
\cite{IsoCosmo,Closed,Bohr} are determined by a wave function
$\psi_{\mu}$ subject to a difference equation of the type
\begin{eqnarray} \label{DiffGen}
&&    (V_{\mu+5}-V_{\mu+3})e^{ik}\psi_{\mu+4}(\phi)- (2+k\gamma^2)
(V_{\mu+1}-V_{\mu-1})\psi_{\mu}(\phi)\\\nonumber
&&+    (V_{\mu-3}-V_{\mu-5})e^{-ik}\psi_{\mu-4}(\phi)
  = -\frac{4}{3}\pi
G\gamma^{-3}\ell_{\rm P}^2\hat{H}_{\rm matter}(\mu)\psi_{\mu}(\phi)
\end{eqnarray}
in terms of volume eigenvalues $V_{\mu}=(\gamma\ell_{\rm
P}^2|\mu|/6)^{3/2}$ \cite{cosmoII} which also show the geometrical
meaning of the label $\mu\in{\mathbb R}$.  A precise model is
specified by the choice of a matter field $\phi$ and Hamiltonian
$\hat{H}_{\rm matter}(\mu)$ and by fixing $k$ to be zero for a flat
model or one for a model of positive spatial curvature. The parameter
$\gamma>0$ is the Barbero--Immirzi parameter of loop quantum gravity
\cite{AshVarReell,Immirzi}, which fixes the scale of spatial
discreteness. There are other versions of the equation, for instance
obtained after a symmetrization procedure which changes the
coefficients but not their large $\mu$ behavior. Our analysis will
also apply to this equation. However, sometimes equations of higher
order appear for which a similar analysis will be more complicated, as
commented on later in this paper. Similarly, partial difference
equations obtained for anisotropic models
\cite{HomCosmo} require more care.

At small $\mu$ the difference equation can be used to understand
quantum aspects of the classical singularity at $\mu=0$
\cite{Sing,DynIn,Essay}, while at large $|\mu|$ one expects
semiclassical behavior \cite{SemiClass}. Here, we are interested in
asymptotic properties of solutions at large $|\mu|$, in particular
their boundedness or oscillatory properties. For simplicity, we base
the analysis on phenomenological matter systems which do not require
an additional matter field. In those cases, the equation takes the
asymptotic form
\begin{equation} \label{asdiff}
 \hat{s}_{\mu+4}-(2+k\gamma^2)\hat{s}_{\mu}+\hat{s}_{\mu-4}=
-\frac{8\pi}{3}G\gamma^2a(\mu)^{-1}H(a(\mu)) \hat{s}_{\mu}
\end{equation}
with $\hat{s}_{\mu}:=e^{ik\mu/4}\psi_{\mu}$ and $a(\mu)^2:=\gamma\ell_{\rm
  P}^2|\mu|/6$. Matter Hamiltonians of interest are the cosmological
constant term $H(a)=\Lambda a^3$, dust with $H(a)$ constant, radiation
$H(a)\propto a^{-1}$, or a free massless scalar with plane wave
initial conditions where $H(a)\propto a^{-3}$. In all these cases, the
right hand side of (\ref{asdiff}) is of the form $-\Lambda |\mu|^v$
where $v$ is constant in a given model, but can in general take both
signs. We simply denote constants characterizing the matter content by
$\Lambda$, even if they do not correspond to a cosmological
constant. Furthermore, without loss of generality we drop the constant
$k$ and define $\gamma$ to be zero in flat models (after having
absorbed it in $\Lambda$ on the right hand side). This brings us to
the general equation
\begin{equation} \label{realmu}
\hat{s}_{\mu+4}-(2+\gamma^2)\hat{s}_{\mu}+\hat{s}_{\mu-4}=
-\Lambda|\mu|^v \hat{s}_{\mu}
\end{equation}
to be studied.

Since the step size of the difference equation is four, defined for a
wave function on the real line, values of $\psi_{\mu}$ between any
$\mu$ and $\mu+4$ are unrelated by the equation. Depending on the
initial conditions, there can thus be oscillations on arbitrarily
small scales, which for intuitive reasons one would like to be
suppressed in semiclassical regimes. If this is the case, a solution
is called pre-classical \cite{DynIn,FundamentalDisc}. This is
partially a matter of initial conditions, but even within a given
sequence at an integer lattice $\mu+4{\mathbb Z}$ can small scale
oscillations develop. Whether or not this can be suppressed depends
now on the difference equation and thus has to be decided by an
asymptotic analysis. Oscillations on larger scales are allowed and
show the presence of a classically allowed regime. Also the occurrence
of these oscillations can be seen by a general analysis.

In regimes where there are no large scale oscillations, which then are
classically forbidden, one expects exponential behavior of
solutions. Generically, the exponentially growing contribution will
dominate such that the issue of boundedness of solutions arises in
particular in models showing a classical recollapse \cite{ClosedExp}.

Both issues, pre-classicality and boundedness, have been used for an
understanding of the semiclassical limit \cite{ClosedExp} and the physical
inner product \cite{IsoSpinFoam}, even though the precise status of
the conditions is not clear. Nevertheless, in those investigations one
often needs to find special solutions, such as bounded ones, which are
difficult to detect numerically in a reliable manner. It is thus
necessary to have analytical tools available. For oscillations,
generating function techniques have been used
\cite{GenFunc,GenFuncBI}, but this is applicable only in special cases
(such as integer $v$ in our equation, and with ignoring the absolute
value signs). The methods used here are more general and also apply to
the boundedness issue.

It may happen that additional conditions will force $\mu$ to take
discrete values \cite{Velhinho} (enumerated by but not necessarily
identical to integer numbers). It is therefore reasonable to
construct tools in such a way that this possible discretization
will be built in. To do so, we first notice that any real number
can be represented as $\mu=n+\delta$ with $n\in \mathbb Z$ and $0 \leq
\delta < 1$, and define $s^{\delta}_n \equiv
\hat{s}_{n+\delta}=\hat{s}_{\mu}$.  Next, we find that the set of
integer numbers splits into four disjoint subsets: ${\mathbb Z}=N_1
\cup N_2 \cup N_3 \cup N_4$ where $N_i=4{\mathbb Z}+i$.
(We could have reduced solutions to only one
subset by allowing $0\leq\delta<4$. However, higher order equations,
such as order four here, naturally arise in loop quantum cosmology,
and so we leave this explicit for the sake of generality.)

When discretization is implied, the delta parameter is being fixed and
the subset concept is very useful. If not, every subset gives rise to
a channel $\hat{N}_i=N_i+\delta$ with $\delta \in [0,1)$. These four
channels cover $\mathbb R$ completely. Equation (\ref{realmu}) can
then be rewritten:
\begin{displaymath}
s^{\delta}_{n+4}-(2+\gamma^2)s^{\delta}_n+s^{\delta}_{n-4}=-\Lambda|n+\delta|^v
s^{\delta}_n\,.
\end{displaymath}
For simplicity, we will drop the label $\delta$ in $s^{\delta}_n$ and
simply write $s_n$. Different sequences will be distinguished from
each other by the appearance of $\delta$ inside the absolute value in
\begin{equation}\label{main}
s_{n+4}-(2+\gamma^2)s_n+s_{n-4}=-\Lambda |n+\delta|^v s_n\,.
\end{equation}

\section{Cosmological constant}

We start with the case of a cosmological constant, i.e.\ $v=1$, in
order to illustrate the main techniques. So equation (\ref{main})
takes the form:
\begin{equation} \label{start}
s_{n+4}-(2+\gamma^2)s_n+s_{n-4}=-\Lambda |n+\delta|s_n
\end{equation}

\subsection{Generic asymptotic behavior}
We define
\begin{displaymath}
\bar{s}_n=(-1)^{\frac{n-q}{4}}s_n
\end{displaymath}
where $n=4m+q$ with $q$ the subset index ($q=i-1$ for $N_i$) and $m$ an
arbitrary integer.  Introducing this into (\ref{start}) results in
\begin{equation}\label{startbar}
\bar{s}_{n+4}+2\bar{s}_n+\bar{s}_{n-4}=\Lambda|n+\delta|\bar{s}_n
\end{equation}
which is more convenient to investigate, starting with the $n\gg1$ region.

To establish the asymptotic behavior we define
\begin{displaymath}
\bar{s}_{n+4}=:h(n+4)\bar{s}_n\,.
\end{displaymath}
We have
\begin{equation} \label{hrec}
h(n+4)=\Lambda |n+\delta|-2-\frac{1}{h(n)}
\end{equation}
which allows expressing $h(n+4)$ in terms of $h(n)$. 

For $n<0$, we can still keep $n$ and $\delta$
positive ($n\gg\Lambda^{-1}$) but substitute $n \to -n$ and $\delta
\to -\delta$ in (\ref{start}). It is then more convenient to use
$\bar{s}_{-(n+4)}/\bar{s}_{-n}$ instead, in order to investigate the
regime of large negative $n$. This ratio is equal to
\begin{displaymath}
\frac{\bar{s}_{-(n+4)}}{\bar{s}_{-n}}=\frac{1}{h(-n)}
\end{displaymath}
and the recurrence now is
\begin{displaymath}
h(-n+4)+\frac{1}{h(-n)}= \Lambda |n+\delta|-2\,.
\end{displaymath}

\subsubsection{Continued fraction}

With (\ref{hrec}) and the analogous equation for $h(n)$, $h(n)=\Lambda
|n+\delta-4|-2-h(n-4)^{-1}$, we rewrite
\begin{displaymath}
h(n+4)=\Lambda |n+\delta|-2-\frac{1}{\Lambda
|n+\delta-4|-2-\displaystyle\frac{1}{h(n-4)}} \,.
\end{displaymath}
This can be iterated, for which it is convenient to use a suitable
notation to indicate that the $k$-th step of iteration has been
performed:
\begin{eqnarray}
h_0 (n+4) &=& \Lambda |n+\delta|-2-\frac{1}{h(n)}\nonumber\\
h_1 (n+4) &=& \Lambda |n+\delta|-2-\frac{1}{\Lambda |n+\delta-4|
-2-\displaystyle\frac{\strut 1}{h(n-4)}}\nonumber\\ 
h_2 (n+4) &=& \Lambda |n+\delta|-2-\frac{1}{\Lambda |n+\delta-4|-
2-\displaystyle\frac{\strut 1}{\Lambda |n+\delta-8|-2 -\displaystyle\frac{
\strut 1}{h(n-8)}}}\nonumber\\\nonumber\\
&\vdots& \nonumber\\\nonumber\\
h_k (n+4) &=& \Lambda |n+\delta|-2 - \frac{1}{\Lambda
|n+\delta-4|-2-\displaystyle\frac{\strut
1}{\Lambda|n+\delta-8|-2-\cdots}}  \label{hp} \\
&&
\qquad \qquad \qquad \qquad \qquad \qquad \qquad \qquad \cdots-\
\frac{1}{\Lambda 
|n+\delta-4k|-2-\displaystyle\frac{\strut 1}{ h(n-4k)}}\nonumber 
\end{eqnarray}
By construction,
\begin{displaymath}
h_k (n+4)= h(n+4) \qquad\mbox{  for all } \quad k=0,1,2,3, \ldots\,.
\end{displaymath}
As we observe, in the $k$-th step we have $k$ denominators with
expressions as $\Lambda|\cdot|-2$ and an ending term $-h(n-4k)^{-1}$.
This process of iteration allows investigating the behavior of
$h(n+4)$ in terms of arbitrary initial conditions of the form
$s_{m_0}$, $s_{m_0+4}$.

For $n<0$ we perform the first step of iteration by
\begin{displaymath}
\frac{1}{h(-n)}=\Lambda|n+\delta|-2 -\frac{1}{\Lambda |n+\delta-4|-2
-h(-n+8)}
\end{displaymath}
and have in the $k$-th step
\begin{eqnarray} \label{hn}
\frac{1}{h^{(k)}(-n)}=\Lambda|n+\delta|-2 -\frac{1}{\Lambda
|n+\delta-4|-2 -\cdots}&&\\
\qquad \qquad \qquad \qquad 
&&\cdots -\frac{1}{\Lambda|n+\delta-4k|-2-h(-n+8(k+1))} \nonumber
\end{eqnarray}
with $k=1,2,\ldots$.

\subsubsection{Convergence}

The iteration process leads easily to asymptotic relations (large
$|n|$ behavior). In this section we will elucidate why this is the
case, discussing only $n>0$ such that $n=4m$, $m\gg1$. Then the ending
term for the $k$-th iteration is $-h(4(m-k))^{-1}$. Also for simplicity
we set $k=m$, and thus the two last denominators of $h_k(n+4)$ are
\begin{eqnarray*}
\frac{1}{\Lambda|n+\delta-4(k-1)|-2-
\displaystyle\frac{\strut 1}{\Lambda|n+\delta-4k|-2-
\displaystyle\frac{\strut 1}{h(n-4k)}}}=&&\\
&&\frac{1}{(4+\delta)\Lambda-2+ \displaystyle\frac{\strut 1}{2-\Lambda\delta+
\displaystyle\frac{\strut 1}{h(0)}}}
\end{eqnarray*}
To convince ourselves that choosing $h(0)$ arbitrarily does not
greatly affect the large $n$ behavior of $h(n+4)$, we can simply
reverse the whole procedure. We pick any value of $h(0)$ and, after a
few steps, the system looses any information about it. To see this we
present some examples. Because we always have inverse powers of
$h(n)$, such as in
\begin{displaymath}
h(4)=\Lambda \delta-2- \frac{1}{h(0)}
\end{displaymath}
and 
\begin{displaymath}
h(8)=(4+\delta)\Lambda-2 - \frac{1}{h(4)}\,,
\end{displaymath}
initial values
for $h(0)$ such that
$h(4) \sim 0$
(but not equal) could give rise to singular behavior. Indeed,
$h(8)$ is large, but then
\begin{displaymath}
h(12)=(8+\delta)\Lambda-2 - \frac{1}{h(8)} \simeq (8+\delta)\Lambda-2\,.
\end{displaymath}
So all information about the potentially dangerous value of $h(0)$ has
been lost and the iteration proceeds as in the generic case.

For completeness, let us see now what happens when $h(4)$ is equal to
zero. To proceed we introduce a parameter $\epsilon$ and a new initial
value $h_{\epsilon} (0)$ such that
\begin{displaymath}
\lim_{\epsilon \to 0} h_{\epsilon} (4)= \Lambda \delta-2-
\frac{1}{h_{\epsilon} (0)}=0 \quad \mbox{ and } \quad
h_{\epsilon}(4)\neq 0 
\mbox{ for }\epsilon\not=0\,.
\end{displaymath}
As $\epsilon$ tends to zero, we will find $h(8)$ explode to infinity,
and $h(12)$ tend to $8\Lambda-2$. So this is just the limiting case of
the previous one.


There exists, however, one very special choice of $h(0)$ whose value
is influential on the whole sequence $h(n)$. If we pick $h(0)$
such that the following sequence of an infinite number of constraints
holds,
\begin{equation} \label{seq}
\frac{1}{h(4m)}\simeq (4m+\delta)\Lambda-2 \qquad \mbox{ for all
}m\in{\mathbb N}\,,
\end{equation}
then cancellations occur within the continued fraction step after step
such that the previously observed ``forgetfulness'' cannot be
realized.

First of all, we can see that if the above sequence of constraints
holds, it will hold only for one, unique initial value which we call
$h_{\rm bounded} (0)$ (unique for given $\Lambda$, $\delta$ and
$k$). In App.~\ref{A:hbound} we present more discussion on this choice
and prove that it leads to a vanishing solution for large $n$. The
discussion above and in App.~\ref{A:hbound} signals that bounded
solutions are permitted by our asymptotic relations. Furthermore, we
will present tools, which allow us to find bounded solutions in a
simple and explicit manner.

For values of $h(0)$ slightly different from $h_{\rm bounded}(0)$, the
iteration undergoes locally abrupt behavior, but after a few steps it
smoothes again. So even for $h(0)$ very close to $h_{\rm bounded} (0)$
the information about the initial value is being lost. 

If $h(0)$ is different from $h_{\rm bounded}(0)$ it is useful to split
the fraction (\ref{hp}) into two pieces (recall that
$k=m=\frac{n}{4}$). The lower part of the fraction, which is affected
by initial value choices for which possibly the $\Lambda|\cdot|$ term
does not dominate, will be called ``part B''. The upper part,
unaffected by the initial value choice and for which
$\Lambda|\cdot|-2$ is much bigger than one, will be called ``part
A''. Thus, there is a $k_{\rm B}<k$ such that part B is identical to
$h_{k_{\rm B}}(n+4)$, and part A is $h_{k-k_{\rm B}}(n+4)$ with part B
used as initial condition.

{}From our arguments, we conclude that part B will become a small
initial value contributing to part A. So what is left is part A, which
is convergent but gives a negligible contribution, compared to the
leading $\Lambda|n|$, as we move to the large $n$ regime. The same
reasoning applies for any choice of an initial value
$h(n-4k)$. It is also easy to convince oneself that the above
arguments are independent of our previous choice $n=4m$.


\subsubsection{Generic growth}

If the initial value is different from $h_{\rm bounded}$, there are at
most finitely many near cancellations, and the following asymptotic
relations (for large $n$) hold
\begin{displaymath}
\frac{s_{n+4}}{s_n}=-h(n+4)\simeq 2-\Lambda|n|
\end{displaymath}
\begin{displaymath}
\frac{s_{-(n+4)}}{s_{-n}}=-\frac{1}{h(-n)} \simeq 2-\Lambda |n|
\end{displaymath}
since the fraction part of (\ref{hp}) or (\ref{hn}) can be ignored.

\subsection{Pre-classicality}

The previous results allow us to draw conclusions for the
pre-classicality of solutions, in particular alternating behavior.

\subsubsection{Subset splitting}

The subset splitting presented in Sec.~\ref{lqc} allows us to reduce
\begin{displaymath}
\hat{s}_{n+4}-2\hat{s}_n+\hat{s}_{n-4}=-\Lambda |n| \hat{s}_n
\end{displaymath}
to four equations
\begin{displaymath}
 s_{4(m+1)}-2s_{4m}+s_{4(m-1)}=-\Lambda |4m+i+\delta| s_{4m}
\end{displaymath}
each valid on one set $N_i$ ($i=1,2,3,4$).
Each of them is of second order, and thus needs two initial
values.  There are no mixing terms, i.e.\ terms that mix $s_n$ from
different $N_i$ subsets. So these four equations are fully
independent.

Because for large $n$ (and for ratios different from $h_{\rm bounded}$)
\begin{displaymath}
\frac{s_{n+4}}{s_n}\simeq 2-\Lambda|n|
\end{displaymath}
and analogously for $s_{-(n+4)}/s_{-n}$, we see that the asymptotic
behavior excludes solutions where $s_n$ and $s_{n+4}$ (also $s_{-n}$
and $s_{-(n+4)}$) are of the same sign. This means that
pre-classicality within one subset $N_i$ is not
preserved. Additionally, it may happen that the solution between
different subsets is not pre-classical. Because, as we have seen,
the original equation reduces to four independent equations of second
order, for each equation we need two initial values. It may then
happen, for example, that for $n \in N_1$ $s_n$ oscillates with
different phase than for $n \in N_2$.

Since we have just proved that generic solutions of (\ref{start}) are
not pre-classical within a subset, non-preclassicality between
different subsets is only of secondary interest. However, we will meet
equations which are pre-classical within a subset, and therefore it
would be of interest to remove non-preclassical behavior between
different subsets on the whole set $\mathbb Z =N_1
\cup N_2 \cup N_3 \cup N_4$.

A simple example of such an equation is (\ref{startbar}). The same
reasoning as above implies now that $\bar{s}_{n+4}$ and $\bar{s}_{n}$
(also $\bar{s}_{-(n+4)}$ and $\bar{s}_{-n}$) are of the same sign, so
the solutions are pre-classical within a subset. We will show how to
remove non-preclassical behavior among different subsets by a proper
choice of initial values. In the next part we will present
generalization to other models.

\subsubsection{Determining regimes leading to solutions of constant sign}

Again, we start with the positive $n$ direction of iteration which
will simply generalize to the negative one. We rewrite
(\ref{startbar}) in a form
\begin{equation}\label{rewrite}
\bar{s}_{n+4}=a(n)\bar{s}_n-\bar{s}_{n-4}
\end{equation}
where
\begin{displaymath}
a(n)=\Lambda |n+\delta|-2\,.
\end{displaymath}

Shifting the label by four and substituting $\bar{s}_n$ in the right
hand side of (\ref{rewrite}), we get $\bar{s}_{n+4}$ in terms of
$\bar{s}_{n-4}$ and $\bar{s}_{n-8}$. After $k$ iteration steps we
reach
\begin{equation}\label{ansatz}
\bar{s}^{(k)} _{n+4}=f^{(k)} (n)\bar{s}_{n-4k}-g^{(k)} (n)\bar{s}_{n-4(k+1)}
\end{equation}
where $f$ and $g$ are functions of $n$ whose structure is of
course $k$-dependent.
For step $k+1$ we have
\begin{displaymath}
\bar{s}^{(k+1)} _{n+4}=f^{(k+1)} (n)\bar{s}_{n-4(k+1)}- g^{(k+1)} (n)
\bar{s}_{n-4(k+2)}
\end{displaymath}
and from (\ref{rewrite})
\begin{displaymath}
\bar{s}_{n-4k}=a(n-4(k+1))\bar{s}_{n-4(k+1)}-\bar{s}_{n-4(k+2)}\,.
\end{displaymath}
Substituting this in (\ref{ansatz}) we obtain
\begin{displaymath}
\bar{s}^{(k+1)}_{n+4}=(a(n-4(k+1))f^{(k)} (n)-g^{(k)} (n))
\bar{s}_{n-4(k+1)}-f^{(k)} (n)\bar{s}_{n-4(k+2)} \,.
\end{displaymath}
Comparing with previous expressions for $\bar{s}^{(k+1)}_{n+4}$ we
obtain the recurrence relations
\begin{eqnarray} \label{rec1}
f^{(k+1)} (n) &=& a(n-4(k+1)) f^{(k)} (n) - f^{(k-1)} (n)\\
g^{(k+1)} (n) &=& f^{(k)} (n)\label{rec2}
\end{eqnarray}
having already used relation (\ref{rec2}) in (\ref{rec1}).
To close these recurrence relations, we notice that
\begin{displaymath}
f^{(0)}(n)=a(n) \quad \mbox{ and }\quad  g^{(0)}(n)=f^{(-1)} (n)=1\,.
\end{displaymath}

We will now focus on relation (\ref{rec1}), which has a symmetric form
similar to (\ref{start}). This suggests to introduce
\begin{displaymath}
\rho (k+1,n)=\frac{f^{(k+1)} (n)}{f^{(k)} (n)}
\end{displaymath}
and rewrite equation (\ref{rec1}) as
\begin{equation} \label{rho}
\rho(k+1,n)=a(n-4(k+1))-\frac{1}{\rho(k,n)}\,.
\end{equation}
If we choose to use as initial values $\bar{s}_{4m_0+q}$ and
$\bar{s}_{4(m_0-1)+q}$ ($q$ is the subset constant for $N_i$, defined
by $q=i-1$, and $m_0$ indicates the initial values in a given subset),
we may write
\begin{equation} \label{solution}
\bar{s}_{n+4}=f^{(\bar{k})}(n)\bar{s}_{4m_0+q}-f^{(\bar{k}-1)}(n)
\bar{s}_{4(m_0-1)+q}
\end{equation}
where $n-4\bar{k}=4m_0+q$ has been substituted by
$\bar{k}=\frac{1}{4}(n-4m_0-q)$. To assure that, for example, the
solution is positive for large values of $n$, the condition
$\bar{s}_{n+4}>0$
easily implies
\begin{displaymath}
\sgn(f)(s_{4m_0+q}-r(\bar{k})s_{4(m_0-1)+q})>0
\end{displaymath}
with
\begin{displaymath}
r(\bar{k})\equiv r(m_0,q)=\lim_{n \to \infty} \frac{1}{\rho(\bar{k},n)}
\end{displaymath}
and $\sgn(f)$ is the sign of the $f$ coefficient which is discussed in
App.~\ref{A:fsign}.

To find $r(m_0,q)$ it is sufficient to use equation
(\ref{rho}). Arguments on convergence of (\ref{rho}) are given in
App.~\ref{A:rexist}.  So by a proper initial value choice we can make
all four solutions on $\mathbb Z$ positive or negative. If $\mu$ is
not discrete, one proceeds as before and makes all solutions positive
or negative for all $\delta \in [0,1)$.

Equation (\ref{rho}) is easy to handle, but nevertheless it is
instructive to discover some properties of this equation for a few
particular cases (for simplicity putting $\delta=0$).

\paragraph{1.} Subset $N_1$, $\bar{s}_0$ and $\bar{s}_{-4}$ as initial
values:

Then $\bar{k}=\frac{n}{4}$ and shifting variables in (\ref{rho}) to
\begin{equation} \label{rhocont}
\rho(k,n)=a(n-4k)-\frac{1}{a(n-4(k-1))-
\displaystyle\frac{\strut 1}{a(n-4(k-2))-\cdots}}
\end{equation}
we obtain
\begin{eqnarray*}
\rho(n/4,n)=-2-\frac{1}{4\Lambda-2-
\displaystyle\frac{\strut 1}{8\Lambda-2-\cdots}}&&\\
&&\cdots-\frac{1}{n\Lambda-2}\,.
\end{eqnarray*}
As expected, this expression is for large $n$ almost independent of
$n$. The only parameter left is $\Lambda$, of which two values are
particularly interesting:

a)  $\Lambda = 0$:
Because
\begin{displaymath}
\frac{1}{-2- \displaystyle\frac{\strut 1}{-2- \cdots}}
=-1
\end{displaymath}
we get
\begin{equation}
 \lim_{n\to \infty}\frac{1}{\rho(n/4,n)}=-1
\end{equation}
in agreement with the results in App.~\ref{A:atwo}.

b)  $\Lambda\gg1$:
On the right hand side of Eq.~(\ref{rhocont}) the fraction can then be
neglected, and we have
\begin{displaymath}
\lim_{n \to \infty} \frac{1}{\rho(n/4,n)}=-\frac{1}{2}\,.
\end{displaymath}
This subset and these initial values are, however, very special
because for $k=\bar{k}=\frac{n}{4}$ the $\Lambda$-term in $a(n-4k)$ in
(\ref{rhocont}) does not contribute. In other subsets, this does not
happen.

\paragraph{2.} Subset $N_2$, $\bar{s}_1$ and $\bar{s}_{-3}$ as initial values:

This time, $n-4\bar{k}=1$ such that $\bar{k}=\frac{n-1}{4}$ and
\begin{equation}\label{rholambda}
\rho(\bar{k},n)=\Lambda-2-\frac{1}{5\Lambda-2-
\displaystyle\frac{\strut 1}{9\Lambda-2-\cdots}}
\end{equation}
We observe, as anticipated, the occurrence of $\Lambda$ at leading
order. This follows
from the subset splitting where for subset $N_1$ we have
\begin{displaymath}
\bar{s}_{4(m+1)}+2\bar{s}_{4m}+\bar{s}_{4(m-1)}=\Lambda |4m| \bar{s}_{4m}\,.
\end{displaymath}
If we put $m=0$ in the above equation we see that the recurrence starts with
$\bar{s}_4=-2\bar{s}_0-\bar{s}_{-4}$ which is $\Lambda$-independent.
For $N_2$, however, we have
\begin{displaymath}
\bar{s}_{4(m+1)+1}+2\bar{s}_{4m+1}+\bar{s}_{4(m-1)+1}= \Lambda |4m+1|
\bar{s}_{4m+1}
\end{displaymath}
and putting $m=0$ leads to
$\bar{s}_5=(\Lambda-2)\bar{s}_1-\bar{s}_{-3}$.  So $\Lambda$ appears
at the zeroth iteration step. Now it is essential that due to
(\ref{rholambda}) the regime for constant sign is sensitive to values
of $\Lambda$ in contrast to the subset $N_1$. It is clear that this is
the generic behavior, $N_1$ being the only exception. \\

For negative $n$ we proceed similarly, starting with
$\bar{s}^{(k)}_{-(n+4)}=\bar{f}^{(k)}(n)\bar{s}_{-n+4k}-
\bar{g}^{(k)}(n)\bar{s}_{-n+4(k+1)}$
such that
\begin{eqnarray*}
\bar{f}^{(k+1)} (n) &=& a(n-4(k+1))\bar{f}^{(k)}(n)-\bar{f}^{(k-1)} (n)\\
\bar{g}^{(k+1)} (n) &=& \bar{f}^{(k)} (n)
\end{eqnarray*}
allows us to write
$\bar{s}_{-(n+4)}=\bar{f}^{(\bar{k}-1)}(n)\bar{s}_{4(m_0-1)+q}-
\bar{f}^{(k)}(n)\bar{s}_{4m_0+q}$
with $\bar{k}=\frac{1}{4}(n+4m_0+q)$.

\subsection{Properties of initial values}

In this subsection we assume positive $n$; the generalization to
negative $n$ is immediate. 

\subsubsection{Asymptotic roles}

Using $\rho$ we write
\begin{equation}\label{invarep}
\bar{s}_{n+4}=f^{(\bar{k})}(n)\left(\bar{s}_{4m_0+q}-
\frac{1}{\rho(\bar{k},n)}\bar{s}_{4(m_0-1)+q}\right)
\end{equation}
and, using the $n$-independence of $\rho$ for large $n$, clearly see
the different roles played by the two initial values. The leading
order contribution to $\rho$ for an arbitrary subset and arbitrary
value of $m_0$ is
\begin{displaymath}
\rho(\bar{k},n)\simeq \Lambda|4m_0+q+\delta|-2
\end{displaymath}
and thus it can happen that $\rho(\bar{k},n)^{-1}\sim 0$ is a small
number, depending on where initial conditions are stated ($m_0$) in
relation to $1/\Lambda$. From Eq.~(\ref{invarep}) we conclude that the
large $n$ behavior of $\bar{s}_{n+4}$ then depends very weakly on
$\bar{s}_{4(m_0-1)+q}$. Thus, $\bar{s}_{4m_0+q}$ dictates the rate of
growth, which can easily be checked numerically as illustrated in
Tab.~\ref{SlowDep}.

\begin{table}
\begin{center}
 \begin{tabular}{c|cccc|cccc}
 $\bar{s}_5$ & 0 & 1 & 2 & 10 & 5 & 5 & 5 & 5\\
$\bar{s}_9$ & 5 & 5 & 5 & 5 & 0 & 1 & 2 & 10\\\hline
$10^{-130}\bar{s}_{205}$ & $\strut 147$ & $148$ &
$148$ & $154$ & $\strut 3.42$ & $32.8$ & $62.2$ & $297$
\end{tabular}
\end{center}
\caption{Dependence on initial values in the subset $N_2$ with
$\Lambda=5$, $\delta=0$, $m_0=2$. \label{SlowDep}}
\end{table}


\subsubsection{Bounded solutions}

{}From the asymptotic relation it follows that the solutions are
generically growing fast and in general are not bounded. However,
using the fact (App.~\ref{A:rexist}) that for $a(n)$ given in
(\ref{rewrite}) the limit $r(m_0,q)$ exists we conclude that special
solutions exist which are bounded. These are those solutions for which
\begin{equation}
\lim_{n\to \infty}\left(\bar{s}_{4m_0+q}-
\frac{1}{\rho(\bar{k},n)}\bar{s}_{4(m_0-1)+q}\right)=0
\end{equation}
i.e.\
\begin{displaymath}
\bar{s}_{4m_0+q}=r(m_0,q)\bar{s}_{4(m_0-1)+q}\,.
\end{displaymath}
We easily see that $r(m_0,q)=h_{\rm bounded}$. This shows that the
infinitely many conditions (\ref{seq}) can indeed be satisfied and
that the value $h_{\rm bounded}$ exists. Most importantly, we now have
a constructive procedure to compute the value for any given model via
$r(m_0,q)$.

If one puts $\bar{s}_{4(m_0-1)+q}=0$ and compares this with (\ref{solution})
and the asymptotic relations one concludes that $f^{(\bar{k}(n))}(n)$
tends to (plus or minus) infinity and so is responsible for the
asymptotic behavior. Putting $\bar{s}_{4m_0+q}=0$ one finds that the
same must hold for $f^{(\bar{k}(n)-1)}(n)$. Looking once again at
(\ref{solution}) allows to understand why the solutions are bounded
for $h_{\rm bounded}$ and why this value is unique (for given
$\Lambda$, $\delta$ and $m_0$). Only then is the growth of
$\bar{s}_{4m_0+q}f^{(\bar{k})}(n)$ canceled by the growth of
$\bar{s}_{4(m_0-1)+q}f^{(\bar{k}-1)}(n)$.

\section{Arbitrary matter}

In this part, we investigate a general class of equations
(\ref{main}):
\begin{equation} \label{diffp}
s_{n+4}-b(n)s_n+s_{n-4}=-\Lambda |n+\delta|^v s_n
\end{equation}
with $b(n)=2+\gamma^2$, and real $v$. We must
distinguish two cases.
\subsection{$v>0$}
Here, the generalization from the discussion of a cosmological
constant is straightforward. First, we introduce
\begin{displaymath}
s_{n}=(-1)^{\frac{n-q}{4}}\bar{s}_n
\end{displaymath}
(where $q$ is the subset index). Equation (\ref{diffp}) then has the form
\begin{equation} \label{diffbar}
\bar{s}_{n+4}+b(n)\bar{s}_n+\bar{s}_{n-4}=\Lambda|n+\delta|^v \bar{s}_n
\end{equation}

We start with the positive $n$ direction of iteration and, to obtain the
asymptotic behavior, define again
\begin{displaymath}
h(n+4)=\frac{\bar{s}_{n+4}}{\bar{s}_n}
\end{displaymath}
which, from (\ref{diffbar}), satisfies
\begin{equation}
h(n+4)=\Lambda|n+\delta|^v-2-\gamma^2-\frac{1}{h(n)}\,.
\end{equation}
For the negative $n$ direction of iteration we have
\begin{displaymath}
\frac{\bar{s}_{-(n+4)}}{\bar{s}_{-n}}=\frac{1}{h(-n)}
\end{displaymath}
and
\begin{displaymath}
h(-n+4)+\frac{1}{h(-n)}= \Lambda |n+\delta|^v-2-\gamma^2\,.
\end{displaymath}

The convergence arguments of the preceding section still apply, and we
obtain
\begin{prop} \label{Prop:Bound}
 Let $s_n$ be a solution to the difference equation (\ref{diffp}) with
 $v>0$. If the ratio of initial values is different from $h_{\rm
 bounded}$, then the rate of growth is given by
\begin{equation} 
\frac{s_{n+4}}{s_n}= -\frac{\bar{s}_{n+4}}{\bar{s}_n}=-h(n+4)=
2+\gamma^2-\Lambda |n|^v
\end{equation}
for large positive $n$, and by
\begin{equation}
\frac{s_{-(n+4)}}{s_{-n}}= -\frac{\bar{s}_{-(n+4)}}{\bar{s}_{-n}}=
2+\gamma^2-\Lambda|n|^v 
\end{equation}
for large negative $n$.
\end{prop}

So we conclude again that solutions within a subset are not
pre-classical. Possible non-preclassicality among different subsets
can be removed applying a generalized regime finding method, presented
in Sec.~\ref{InRep}, to each subset equation (\ref{diffbar}).

\subsection{$v<0$}

This model is very interesting physically, because it describes a
universe with matter or radiation. As we will see, it contains many
new features. Let us rewrite equation (\ref{diffp}) in the form
\begin{displaymath}
s_{n+4}-(2+\gamma^2-\Lambda |n+\delta|^v)s_n +s_{n-4}=0
\end{displaymath}
and define
\begin{displaymath}
F(n)=2+\gamma^2-\Lambda |n+\delta|^v\,.
\end{displaymath}
First, for negative $v$ we have $\lim_{n \to \pm \infty} \
F(n)=2+\gamma^2$ so we conclude that the results of Appendix
\ref{A:vn} will be useful, and Eq.~(\ref{diffconst}) there is
indeed an asymptotic equation.

\subsubsection{$\gamma \neq 0$}
\label{gammanotzero}

We use $h(n+4)=s_{n+4}/s_n$ and reach
\begin{displaymath}
h(n+4)=2+\gamma^2-\Lambda|n+\delta|^v-
\frac{1}{2+\gamma^2-\Lambda|n+\delta-4|^v- \displaystyle
\frac{\strut 1}{2+\gamma^2-\Lambda|n+\delta-8|^v-\cdots}} 
\end{displaymath}
and
\begin{displaymath}
\frac{1}{h(-n)}=2+\gamma^2-\Lambda|n+\delta|^v-
\frac{1}{2+\gamma^2-\Lambda|n+\delta-4|^v- \displaystyle \frac{\strut 1}{
2+\gamma^2-\Lambda|n+\delta-8|^v-\cdots}} 
\end{displaymath}
In the large $n$ limit, using results of \ref{A:asmall}, we have
\begin{prop}  \label{asvneg}
 Let $s_n$ be a solution to the difference equation (\ref{diffp}) with
 $v<0$ and $\gamma\not=0$. If the ratio of initial values is different
 from $h_{\rm bounded}$, then the rate of growth is given by
\begin{equation} 
\frac{s_{n+4}}{s_n}=h(n+4)=2+\gamma^2-x
\end{equation}
for large positive $n$, and by
\begin{equation}
\frac{s_{-(n+4)}}{s_{-n}}=\frac{1}{h(-n)}=2+\gamma^2-x
\end{equation}
for large negative $n$, where
\begin{displaymath}
x=\frac{2+\gamma^2-\sqrt{\gamma^4+4\gamma^2}}{2} <1 \,.
\end{displaymath}
\end{prop}
We conclude that solutions are preclassical within a subset,
i.e. $s_{n+4}$ and $s_n$ (also $s_{-(n+4)}$ and $s_{-n}$) are of the
same sign for all large $n$ (this means that also long distance
oscillations are not possible). The only problem is to remove possible
oscillating behavior between subsets, which is an easy task due to
Sec.~\ref{InRep} (the limit of $\rho(\bar{k},n)^{-1}$ will exist) and
is achieved by picking up proper initial values.

However this does not exhaust all properties of the $\gamma \neq 0$
case. Long distance and short distance oscillations are excluded for
large $n$, but they still may occur for small $n$ (the standard of
smallness is parameter dependent).

\begin{figure}
\begin{center}
\includegraphics[scale=1]{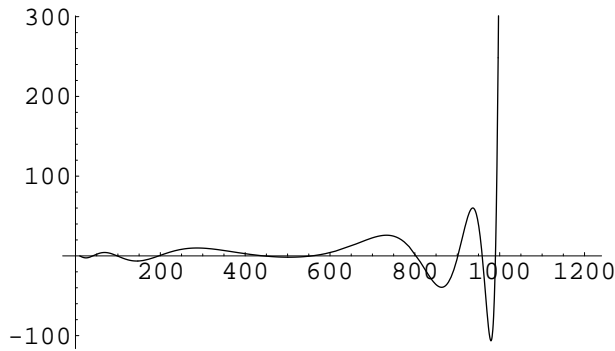}
\end{center}
\caption{Typical behavior of a solution with last oscillation at
$\bar{n}=1000$ and small $m_0$. \label{fig:lastoscill}}
\end{figure}

If one picks $\Lambda$ and $\gamma$ in such a way that $F(n)>2$ for
all $n$, then (with help of Appendix \ref{A:vn}) solutions are always
non-oscillating. But for other choices one in general has $F(n)<2$ for
$n\in [-n_0,n_0]$ and $F(n)>2$ for $n\in {\mathbb Z}\setminus
[-n_0,n_0]$.  As long as $F(n)<2$ the oscillating nature of solutions
may manifest itself which, however, is not guaranteed: The condition
$F(n)=2$ has solution $|\bar{n}|=
(\gamma^2/\Lambda)^{1/v}=(\Lambda/\gamma^2)^{1/|v|}$.  One may call
$\bar{n}$ the value(s) of last oscillation. Depending on the phase of
oscillation in a neighborhood of $\bar{n}$, the solution may lose its
oscillating nature just before or just after $\bar{n}$. It is
instructive to illustrate this feature with a numerical example:
$N_3$, $\Lambda=5$, $\gamma=0.1$, $p=-1$, $s_2=2$, $s_6=3$. The
solution passes zero for the last time at $n=474$ while the value of
last oscillation is $\bar{n}=500$ (see also Fig.~\ref{fig:lastoscill}).

The necessary condition for small $n$ oscillations is now easy to
state: The value of last oscillation must be greater than $4m_0+q$ or
smaller than $4(m_0-1)+q$, at least by a value of four. Otherwise,
oscillations will not even have space to start. A sufficient
condition will be formulated in what follows.

\subsubsection{$\gamma=0$}

This case is profoundly different because $F(n) <2$ for all $n$.
Thus, with results in Appendix \ref{A:vn} we conclude that generic
solutions are oscillating. We start investigations with the $v=-1$
case, moving to the large $n$ regime. For any $\epsilon>0$ we may
always find such $N_0$ that for $N>N_0$ we have $2-\epsilon < F(n) <
2$.  We can use (\ref{sols}) in Appendix \ref{A:vn} as an approximate
solution, with one caveat. Since we are in the large $n$ regime, we
cannot pick arbitrary $n_0$ in (\ref{sols}). Rather, it should be some
value at which the large $n$ regime has already started. This means
that, in general, we do not know anything about $s_{n_0}$,
$s_{n_0-4}$, which are responsible for the amplitude of
oscillation. However, the period of oscillation is independent of
$s_{n_0}$, $s_{n_0-4}$, and we may use (\ref{sols}) to extract the
period of oscillation. Using (\ref{solt}) and (\ref{sols}) we find the
oscillating part of the solution for large $n$ to be given by
$\sin(\omega n/4)$ with period of oscillation $T=8\pi/\omega$.  It is
more convenient to use half of the period,
$\bar{T}:=T/2=4\pi/\omega$ which will be the distance between
two subsequent zeros of oscillation. (Because $n$ is discrete, and its
progress is four, it rarely happens that a solution takes zero
value. Usually we have $s_n$ and $s_{n+4}$ of different sign. By zero
of oscillation we mean then a number $z=n$ or $z=n+4$, depending on
which absolute value is smaller, $s_n$ or $s_{n+4}$.) With
$\omega(a)=\tan^{-1} (\sqrt{4-a^2}/a)$ and substituting $a$ by
$2-\Lambda/|n|$ we reach an approximate half period of oscillation of
\begin{equation} \label{Tbar}
\bar{T}(n)= \frac{4\pi}{\tan^{-1}\left((\sqrt{4\Lambda}{|n|}-
\Lambda^2/|n|^2)/(2-\Lambda/|n|)\right)}\,. 
\end{equation}
With $\tan^{-1} (x)=x-\frac{x^3}{3}+(x^5)$ for very large $n$ we can
write
\begin{equation}
\bar{T}(n)\sim\frac{4\pi}{\sqrt{\Lambda}}\sqrt{|n|}
\end{equation}
Since (\ref{Tbar}) is indeed independent of $s_{n_0}$ and $s_{n_0-4}$,
it should be a good approximation. To compare it with a numerical
iteration we proceed as follows. Let $z_1$ and $z_2$ be two subsequent
zeros of a solution. We define $(z_1+z_2)/2=n$ and use this value in
(\ref{Tbar}) to check if $\bar{T}(n) \simeq |z_1-z_2|$.
This is confronted with a few numerical results in Tab.~\ref{halfper},
showing indeed good agreement.

\begin{table}
\begin{center}
\begin{tabular}{c|cc|ccc|c}
$\Lambda$ & \multicolumn{2}{c|}{initial values} & $z_1$ & $z_2$ &
$z_2-z_1$ & $\bar{T}(n)$\\\hline 1 & $s_0=2$ & $s_4=3$ & 332 & 604 &
272 & 271.83\\ 2 & $s_1=2$ & $s_5=3$ & 973 & 1269 & 296 & 297.48\\ 10
& $s_{10}=2$ & $s_{14}=5$ & 1806 & 1978 & 172 & 172.81\\
\end{tabular}
\end{center}
\caption{Numerical compared to analytical values for the half-period
(\ref{Tbar}). \label{halfper}}
\end{table}


As already mentioned it is not possible to derive an exact value for
the amplitude from (\ref{sols}). However, it is possible to derive the
rate of growth which is proportional to $2(4\Lambda/|n|-
\Lambda^2/|n|^2)^{-1/2} \sim \sqrt{|n|}$.  

Let us now drop the assumption $v=-1$. For large $n$ the oscillating
part of the solution is $\sin(\omega(n)n)$ with
\begin{displaymath}
\omega(n)=\frac{1}{4}\
{\tan^{-1}\left(\frac{\sqrt{4\Lambda/|n|^r-
\Lambda^2/|n|^{2r}}}{2-\Lambda/|n|^r}\right)}
\end{displaymath}
and $r=|v|$.  One concludes that oscillations may be suppressed when
$\lim_{n \to \infty} \omega(n) n=\lim_{n\to \infty} n^{1-\frac{r}{2}}$
exists which implies $r\geq 2$. So for $\gamma=0$ we have
\begin{prop}
For $-2 < v <0$ the solutions are oscillating with increasing period
$\frac{2\pi}{\omega(n)}$. For $v\leq-2$ the phase is decreasing so
rapidly that oscillations are suppressed. The regime finding method
presented in Sec.~\ref{InRep} allows then to remove non-preclassical
behavior among subsets.
\end{prop}

\begin{figure}
\begin{center}
\includegraphics[scale=1]{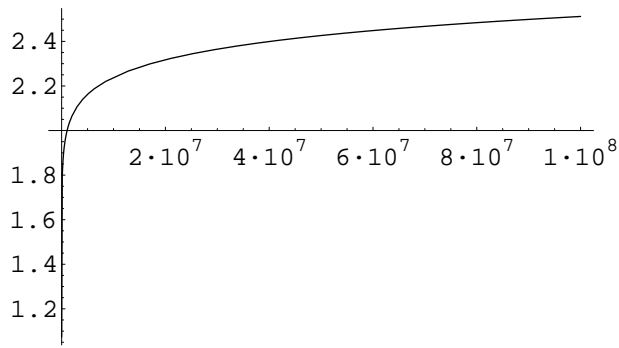}
\end{center}
\caption{Phase $\omega(n)n$ of the approximate solution with $v=-1.9$. \label{fig:phase}}
\end{figure}

The oscillating nature of solutions for $-2<v<0$ does not imply that
they are not preclassical. The period of oscillation increases, and
oscillations are of long distance type. For $v$ close to $-2$, no
oscillations occur for extremely large $n$. Fig.~\ref{fig:phase}
illustrates the rate of growth of $\omega(n) n$ for $v=-1.9$.

Using the results of this subsection, it is now easy to state the
sufficient condition for small $n$ oscillations in
\ref{gammanotzero}. The half period of oscillations for small $n$
should be smaller than the value $\bar{n}-4m_0-q$ for the positive $n$
direction of iteration. The generalization to the negative $n$ direction of
iteration is again straightforward.

\subsection{Generic asymptotic growth}

Solving the difference equation for $s_n$ given by the asymptotic
behavior we obtain a growth of the form $\Lambda^{n/4}\prod_{i=1}^{n/4}
|i|^v$ for $v>0$ from Prop.~\ref{Prop:Bound} and $c^n$ with a constant
$c$ for $v<0$ and $\gamma\not=0$ from Prop.~\ref{asvneg}. For $v<0$
and $\gamma=0$ solutions are oscillatory with growing amplitude at
large $|n|$. We summarize this by

\begin{theo}[Generic asymptotic growth.] \label{theo:GenAs}
 Let $\psi_{\mu}$ be a solution to the difference equation
 (\ref{DiffGen}) without matter field. If $v>0$, then $|\psi_{\mu}|$
 increases stronger than exponentially for $|n|\to\infty$. If $v<0$,
 then $|\psi_{\mu}|$ increases exponentially unless $\gamma=0$ in which
 case the increase in amplitude is given by a power law depending on
 $v$.
\end{theo}

\subsection{Initial value representation}
\label{InRep}

So far, we discussed mainly asymptotic properties where the precise
form of the coefficients in (\ref{DiffGen}) is irrelevant. However,
when we now turn to the initial value representation of bounded
solutions, we must take into account that these initial values may be
stated at small $n$ where the asymptotic form of the coefficients is
not precise.  In order to prove that the initial value representation
is still valid we thus have to use the most general form of our
difference equation
\begin{equation}\label{diffgen}
a(n)s_{n+4}-b(n)s_{n}+c(n)s_{n-4}=f(n)s_n
\end{equation}
where $b(n)$, $c(n)$, $f(n)$ are arbitrary functions, and for $a(n)$
we only assume $a(n) \neq 0$ for all $n$ even though generalizing to
the case where $a(n)$ has a finite number of zeros is not difficult.

Naturally, this equation again splits into four
subsets $N_1, N_2, N_3, N_4$. We make an ansatz
\begin{equation} \label{sk}
s^{(k)}_{n+4}=f^{(k)}(n) s_{n-4k}-g^{(k)}(n) s_{n-4(k+1)}
\end{equation}
such that $s^{(k+1)}_{n+4}=
f^{(k+1)}(n)s_{n-4(k+1)}-g^{(k+1)}(n)s_{n-4(k+2)}$.  On the other hand,
from (\ref{diffgen})
\begin{equation} \label{snk}
s_{n-4k}=\frac{F(n-4(k+1))}{a(n-4(k+1))}s_{n-4(k+1)}-
\frac{c(n-4(k+1))}{a(n-4(k+1))}s_{n-4(k+2)} 
\end{equation}
with $F(n)=b(n)+f(n)$. Defining
\begin{equation} \label{hInRep}
h(n-4(k+1))=\frac{F(n-4(k+1))}{a(n-4(k+1))}
\end{equation}
and 
\begin{equation} \label{jInRep}
j(n-4(k+1))=\frac{c(n-4(k+1))}{a(n-4(k+1))}
\end{equation}
and inserting (\ref{snk}) in (\ref{sk}) using above definitions, we reach
\begin{displaymath}
f^{(k+1)}(n)=h(n-4(k+1)) f^{(k)}(n)-g^{(k)}(n)
\end{displaymath}
\begin{displaymath}
g^{(k+1)}(n)=j(n-4(k+1)) f^{(k)}(n)
\end{displaymath}
and thus
\begin{equation} \label{fInRep}
f^{(k+1)}(n)=h(n-4(k+1)) f^{(k)}(n)-j(n-4k) f^{(k-1)}(n)\,.
\end{equation}
Introduction of
\begin{equation} \label{rhoInRep}
\rho(k,n)=\frac{f^{(k)}(n)}{f^{(k-1)}(n)}
\end{equation}
shows that the initial value representation is still valid, and has the form
\begin{equation}
s_{n+4}=f^{(\bar{k})}(n) \left(s_{4m_0+q}-\frac{j(4m_0+q)}{\rho(\bar{k},n)}
s_{4(m_0-1)+q}\right) 
\end{equation}
for the positive direction of iteration, the negative one following
analogously. This representation is particularly important when the
limit $\lim_{n\to \infty} \rho(\bar{k}(n),n)^{-1}$ exists. Violations
of pre-classicality among different subsets may then be removed and
bounded solutions exist. We have

\begin{theo}[Initial value representation of bounded solutions.] 
\label{theo:Bounded}
 Let $a(n)$, $b(n)$, $c(n)$ and $f(n)$ be functions such that
 $\lim_{n\to \infty} \rho(\bar{k}(n),n)^{-1}=:r(m_0,q)$ exists, where
 $\rho(k,n)$ defined in (\ref{rhoInRep}) depends on the functions
 through (\ref{fInRep}) and (\ref{hInRep}), (\ref{jInRep}) with
 $F(n)=b(n)+f(n)$.

 Then the difference equation (\ref{diffgen}), such as
 (\ref{DiffGen}), in any subset $N_{q+1}$ has a solution bounded at
 positive $n$ whose initial values specified at $m_0$ satisfy
 $s_{4m_0+q}=r(m_0,q)j(4m_0+q)s_{4(m_0-1)+q}$.
\end{theo}

\section{Example: Euclidean cosmological constant model}
\label{Eucl}

For a Euclidean cosmological constant model the equation
\begin{equation}\label{eucl}
s_{n+4}-2s_n+s_{n-4}=\Lambda n s_n
\end{equation}
has been studied \cite{IsoSpinFoam}, which is here of interest because
special solutions are known (note the absence of absolute values
around $n$ in this toy model, which leads to explicitly known
solutions). This can then be compared with our methods which easily
apply to this equation, in particular the regime finding method and
the condition for bounded solutions.

For positive $n\in N_1$, Eq.~(\ref{eucl}) has solution
\begin{equation}
s_{n}=C^{N_1}_1
J_{\frac{n}{4}+\frac{1}{2\Lambda}}\left(\frac{1}{2\Lambda}\right)+ C^{N_1}_2
Y_{\frac{n}{4}+\frac{1}{2\Lambda}}\left(\frac{1}{2\Lambda}\right)\,. 
\end{equation}
For $n \to \infty$ the $Y$ Bessel function tends to minus infinity and
the $J$ Bessel function to zero. Generically, the function $Y$ thus
dominates, which agrees with our results for the generic asymptotic
growth (Theorem \ref{theo:GenAs}): If $v=1$, we have $s_{n+4}\sim
\Lambda n s_n$ from Prop.~\ref{Prop:Bound} such that $s_n\sim
(4\Lambda)^{n/4}(n/4-1)!$. With Stirling's formula $n!=\sqrt{2\pi
n}n^n e^{-n}(1+O(n))$ we obtain $s_n\sim 2\sqrt{2\pi}(n-4)^{-1/2}
(\Lambda (n-4))^{n/4}e^{-n/4+1}$. This can easily be seen to agree
with the asymptotic growth of the Bessel function
$Y_{\nu}(z)\sim-\sqrt{2/\pi\nu}(ez/2\nu)^{-\nu}$ in the case
$z=1/2\Lambda$, $\nu=n/4+z$ which determines the solution to the
difference equation.

If we choose $s_{4m_0}$ and $s_{4(m_0-1)}$ as initial values then
$C^{N_1}_2$ is given by
\begin{equation}
C^{N_1}_2=\frac{s_{4(m_0-1)}J_{m_0-1+\frac{1}{2\Lambda}}}{\det(m_0,\Lambda)}
(a_1-a)
\end{equation}
where $\det(m_0,\Lambda)=J_{m_0+\frac{1}{2\Lambda}}
Y_{m_0-1+\frac{1}{2\Lambda}}- Y_{m_0+\frac{1}{2\Lambda}}
J_{m_0-1+\frac{1}{2\Lambda}}$ and $a_1=J_{m_0+\frac{1}{2\Lambda}
}/J_{m_0-1+\frac{1}{2\Lambda}}$, $a=s_{4m_0}/s_{4(m_0-1)}$.  One can
show that $\det(m_0,\Lambda)=\det(0,\Lambda)>0$ and
$J_{m_0-1+\frac{1}{2\Lambda}}\left(\frac{1}{2\Lambda}\right)>0$ for
all positive $\Lambda$ and $m_0$. So if for example
$s_{4(m_0-1)+q}>0$, then for $a<a_1$ the solution for large $n$ is
negative and for $a>a_1$ it is positive. For $a=a_1$, $C^{N_1}_2=0$
and the solution is bounded. In our previous notation (Theorem
\ref{theo:Bounded}), this means $a_1=r(m_0,q)$ for the bounded
solution.

Comparing the results, we obtain a continued fraction representation
for the ratio of values of Bessel functions,
\[
 \frac{J_{m_0-1+\frac{1}{2\Lambda}} \left(\frac{1}{2\Lambda}\right)}{
J_{m_0+\frac{1}{2\Lambda}} \left(\frac{1}{2\Lambda}\right)}=
2+4\Lambda m_0-\frac{1}{2+4\Lambda(m_0+1)-\displaystyle\frac{\strut
1}{2+4\Lambda(m_0+2)-\cdots}}
\]
which in a similar form (with plus instead of minus signs in front of
the fractions, which lead to modified Bessel functions of the first
kind) can also be found in \cite{HAKMEM}.

Generalization to other subsets and negative $n$ is easy to perform,
and also confirms the presented method.

\section{Conclusions}

The methods introduced here can be used to determine the generic
growth of solutions to a difference equation of type
(\ref{asdiff}). Moreover, even in generically unbounded cases there
are analytical conditions for the presence and values of special
initial conditions giving rise to bounded solutions. These conditions
can be evaluated numerically in a constructive manner, without being
sensitive to numerical instabilities which would hamper any trial and
error approach. We have illustrated this in Sec.~\ref{Eucl} in a
model where explicit solutions to the difference equation are known.

Bounded solutions are often preferred, even though a clear reason is
not known unless a physical inner product is used. One may impose
boundedness as a final condition on allowed solutions, but usually in
quantum cosmology initial conditions are imposed or arise at small
volume. This, in general, does not guarantee that a wave function
evolves to a bounded solution. In fact, bounded solutions are very
special if there are classically forbidden regions at arbitrarily
large volume. In such a case, growing solutions generically
dominate. This is familiar from quantum mechanics, and has been shown
here to be realized similarly at the level of difference equations
directly. We emphasize, however, that the situation is more
complicated with matter degrees of freedom, where the matter
Hamiltonian is an operator and methods similar to those used here
would require continued fractions of operators. Similarly, one has to
deal with operator valued continued fractions in anisotropic models
where now $a(n)$, $b(n)$ and $c(n)$ as used in Sec.~\ref{InRep} are
difference operators on labels other than $n$
\cite{HomCosmo,Spin}. Higher order difference equations can also occur
\cite{IsoCosmo,AmbigConstr}, for which we can still introduce $h(n)$
as before. However, if there are also terms with $s_{n\pm 8}$, the
equation for $h(n)$ obtained after dividing by $s_n$ contains
$s_{n+8}/s_n=h(n+8)h(n+4)$ which makes it more complicated to solve.

There are two different origins of unboundedness of solutions of
difference equations. The first appears in classically forbidden
regions, and is not different from unboundedness in the continuum
limit. This happens if $h(n)>1$, which can come from positive
curvature $\gamma>0$ or negative matter terms such as a negative
cosmological constant. There is also a second possibility, where now
the behavior of solutions to a difference equation is completely
different from that of a differential equation. This happens if
$h(n)<-1$, which we have for a positive cosmological constant. In this
case, the solution is not only unbounded but also has alternating
behavior and is thus not pre-classical. Such a behavior cannot occur
in the continuum limit, and expresses the presence of large
curvature. It should be interpreted as a breakdown of the
minisuperspace approximation where the integrated curvature is the
only curvature quantity. Local curvature is small if the cosmological
constant is small, but the integrated curvature $\Lambda|\mu|$ becomes
large just from large volume. As the analysis here shows, this
behavior is clearly distinguished and easily detectable.

\begin{appendix}
\section*{Appendix}

\section{The choice of $h_{\rm bounded}$}
\label{A:hbound}

Let us assume that the sequence of constraints (\ref{seq}) holds. This
means that if we take a large positive value of $n$, then there exists
a small $\epsilon_1 (n)$ such that
\begin{displaymath}
h(n+4)=\Lambda|n|-2-\frac{1}{h(n)}=\epsilon_1 (n)
\end{displaymath}
If we assume $h(n+4)\geq h(n)$ then $-h(n+4)^{-1}\geq-h(n)^{-1}$ and
\begin{displaymath}
h(n+8)=\Lambda|n+4|-2-\frac{1}{h(n+4)}\geq
\left(\Lambda|n|-2-\frac{1}{h(n)}\right)+4\Lambda=4 \Lambda +\epsilon_1 (n) 
\end{displaymath}
implying in turn
\begin{displaymath}
h(n+12)=\Lambda|n+8|-2-\frac{1}{h(n+8)} \geq \Lambda|n+8|-2-\frac{1}{4
\Lambda+\epsilon_1 (n)} \,.
\end{displaymath}
A cancellation is then blocked by the $4\Lambda$ term, which is in
contradiction to the assumption of cancellations for all $n$. So we
have $h(n+4) < h(n)$ and the sequence is decreasing. One can easily
find out that $h(n)$ with the choice $h_{\rm bounded}$ is much smaller
than one but positive. Together with the previous prove this implies
that solutions are vanishing asymptotically. 

One can convince oneself, that this is the case by noting that the
previous observations imply for large $n$ (and positive solutions)
$0<s_{n}<c^{(n-m)/4} s_m$ for some $m$ and $0<c<1$, and thus
$\lim_{n\to \infty}s_n=0$. For negative solutions we have
$0>s_{n}>c^{(n-m)/4}s_m$ and again conclude $\lim_{n\to \infty} s_n=0$.

The generalization to large negative $n$ is straightforward.

\section{Sign of $f$ coefficients and the existence of regimes}
\label{A:f}

\subsection{Sign of $f$ coefficients}
\label{A:fsign}

By definition, the coefficients $f$ obey the recurrence relation
\begin{displaymath}
f^{(k+1)}(n)=a(n-4(k+1))f^{(k)}(n)-f^{(k-1)}(n)
\end{displaymath}
with $f^{(0)}(n)=a(n)$ and $f^{(-1)} (n)=1$.  We will prove that for
$a(n)\geq 2$ for all $n$ all $f$ coefficients are positive. We have
$f^{(1)}(n)=a(n-4)a(n)-1> a(n)=f^{0}(n)$ noting that
\begin{displaymath}
a(n-4)a(n)-1 \geq 2a(n)-1=a(n)+(a(n)-1)>a(n)
\end{displaymath}
because $a(n)-1\geq 1 > 0$ as assumed. Let us assume now that
$f^{(l)}(n)>f^{(l-1)}(n)$ for $l=0,1,\ldots,m$. Then
\begin{displaymath}
f^{(m+1)}(n)=a(n-4(m+1))f^{(m)}(n)-f^{(m-1)}(n)\geq 2
f^{(m)}(n)-f^{(m-1)}(n)>f^{(m)} (n)
\end{displaymath}
and by induction we finish the prove that $f^{(k)}(n)>f^{(k-1)}(n)$
for all $k$. So the sequence ${f^{(k)}(n)}$ is growing strictly
monotonically and its initial values are positive, implying
$f^{(k)}(n)>0$.

We now turn to the case where
\begin{equation}\label{a}
a(n)<2 \qquad \mbox{ for } \ n\in [n_1,n_2] \qquad \mbox{ and } \qquad
a(n)\geq 2 \qquad \mbox{ for } {\mathbb Z}\setminus [n_1,n_2]
\end{equation}
and consider the positive $n$ direction of iteration first. If we take
large $n$ (fulfilling $n\gg n_2$), then $a(n-4k) \geq 2$ for
$k=0,1,2,\ldots,k_0$ where we have $f^{(k)}(n)>f^{(k-1)}(n)>0$ for
$k=0,1,2,\ldots,k_0$, and so $\rho(k,n)>0$ for $k=0,1,2,\ldots,k_0$\,.
We must then investigate what happens with the sign of $f$ for
$k=k_0+1,\ldots,\bar{k}$ (one should note that $\bar{k}-k_0$ does not
depend on $n$). With help of the already established formula
\begin{equation}\label{rhoagain}
\rho(k+1,n)=a(n-4(k+1))-\frac{1}{\rho(k,n)}
\end{equation}
and $\rho(0,n)=a(n)$ we find $\rho(k_0,n)$. We then use this formula
again to see whether or not
$\rho(k_0+1,n)=f^{(k_0+1)}(n)/f^{(k_0)}(n)$ changes its sign. If not,
$f^{(k_0+1)}(n)$ is also positive. We then investigate
$\rho(k_0+2,n)$, and so on. Every time $\rho(k,n)<0$ we have a change
of sign, until we reach the desired value $\rho(\bar{k},n)$. So one
such iteration procedure allows to find both
$\rho(\bar{k},n)$ and $\sgn(f)=f^{(\bar{k})}(n)/|f^{(\bar{k})}(n)|$.

We will prove in the next subsection that for $a(n)$ as in (\ref{a})
$\lim_{n \to \infty} \rho(\bar{k}(n),n)$ exists, and that
$\rho(\bar{k}(n),n)$ is $n$-independent for large $n$, and so must
equal $\rho(k_0 (n),n)$. So it is sufficient to apply the above
procedure for one value of large $n$.

\subsection{Existence of the regime constant $r(m_0,q)$}
\label{A:rexist}

The problem of existence of $r(m_0,q)=\lim_{n \to \infty}
\rho(\bar{k}(n),n)^{-1}$ is closely related to the existence of
continued fractions. We will show this for the positive $n$ direction
of iteration. From Eq.~(\ref{rhoagain}) we have
\begin{displaymath}
\rho(k,n)=a(n-4k)- \displaystyle\frac{\strut 1}{a(n-4(k-1))-
\displaystyle\frac{\strut 1}{a(n-4(k-2))-\cdots}}\,.
\end{displaymath}
Substituting $\bar{k}(n)=\frac{1}{4}(n-4m_0-q)$ we reach
\begin{equation}\label{rhofrac}
\rho(\bar{k}(n),n)=a(4m_0+q)- \displaystyle\frac{\strut
1}{a(4(m_0+1)+q)- \displaystyle\frac{\strut 1}{a(4(m_0+2)+q)-\cdots}}\,.
\end{equation}
So if $a(n)$ is of the form (\ref{a}), there exists $l_0$ such that
for all $l\geq l_0$
we have $a(4(m_0+l)+q) \geq 2$.
With help of results of \ref{A:cont},
we conclude that (\ref{rhofrac}) is convergent, and for large $n$
almost $n$-independent.

\section{The case $v=0$ as an asymptotic equation}
\label{A:vn}

If $v\leq 0$, we are dealing with a difference equation whose
coefficients are almost constant at large $n$ (or exactly constant for
$v=0$) and which is of the form
\begin{equation} \label{diffconst}
s_{n+4}-as_n+s_{n-4}=0
\end{equation}
with a real parameter $a$. This equation can be analyzed by explicit
solutions, but care must be taken in interpreting the results since
the limiting value $a=2$ for large $n$ and $v<0$ is a critical value
of the difference equation.

\subsection{Solutions}

We start with subset $N_1$ and introduce the generating function
$F(x)=\sum^{\infty} _{m=m_0} s_{4m} x^{m}$.  From equation
(\ref{diffconst}) we get
$(x^{-1}-a+x)F(x)-s_{4m_0}x^{m_0-1}+s_{4(m_0-1)}x^{m_0}=0$ and
\begin{displaymath}
F(x)=\frac{x^{m_0} s_{4m_0}-x^{m_0+1} s_{4(m_0-1)}}{1-ax+x^2}\,.
\end{displaymath}
To find solutions we must expand this generating function as a power
series in $x$. One easily sees that the numerator will only result in
a variable shift, and the main task is to expand
\begin{equation}\label{g}
G(x)=\frac{1}{1-ax+x^2}=\sum^{\infty}_{n=0}t_n x^n\,.
\end{equation}
We will use $1-ax+x^2=(x-x_1)(x-x_2)$ with roots
\begin{equation}\label{roots}
x_1=\frac{a+\sqrt{a^2-4}}{2} \qquad\mbox{ and } \qquad
x_2=\frac{a-\sqrt{a^2-4}}{2}
\end{equation}
which are real if and only if $|a|> 2$. 

\subsubsection{$|a|> 2$}
\label{A:alarge}

For this case, we write
\begin{displaymath}
\frac{1}{(x-x_1)(x-x_2)}=\frac{1}{x_1-x_2}
\left(\frac{1}{x-x_1}-\frac{1}{x-x_2}\right)
\end{displaymath}
and $(x-x_1)^{-1}=-x_1^{-1} (1-x/x_1)^{-1}=-x_1^{-1}
\sum_{k=0}^{\infty} (x/x_1)^k$.  From (\ref{g}) we reach
\begin{equation}\label{t}
t_{n}=\frac{1}{x_1-x_2}\left(x_2^{-n-1}-x_1^{-n-1}\right)=
(a^2-4)^{-1/2} \left(x_2^{-n-1}-x_1^{-n-1}\right)\,.
\end{equation}
For the large $m$ behavior we notice that for $a> 2$ we have $x_1>1$
and $x_2<1$.  For large $n$ the second term in bracket in (\ref{t})
becomes negligible and the first one results in $\lim_{n\to\infty}t_n
= \infty$.

\subsubsection{$|a| < 2$}
\label{A:asmall}

We make an analytic continuation to the principal sheet in (\ref{t})
and prove that the $t_n$ remain real. We see that
$x_1=\frac{1}{2}(a+i\sqrt{4-a^2})= x_2^*$ and write $x_1=|r|
e^{i\omega}$, $x_2=|r| e^{-i\omega}$.  Before going further we notice
$|r|=1$ so that the roots are pure phase. We rewrite equation
(\ref{t}) as
\begin{equation} \label{solt}
t_{n}=-\frac{2}{\sqrt{4-a^2}}\sin(\omega (n+1))\,.
\end{equation}
Thus, for $|a|< 2$ solutions are oscillating with constant amplitude
and $a=a_0=2$ is a critical point.

{}From $\omega=\tan^{-1}(\sqrt{4-a^2}/a)$ we find the oscillation period
$T=2\pi/\omega$.  Since $\lim_{a \to 2-}T(a)=\infty$ the closer $a$ is
to two, the longer the period is, and solutions slowly acquire the
non-oscillating nature of $|a|\geq 2$ (in \ref{A:atwo} we show that
also the case $a=a_0=2$ is non-oscillating).

Making the already mentioned variable shift in $F(x)$ we reach
\begin{displaymath}
s_{4m}=s_{4m_0}t_{m-m_0}-s_{4(m_0-1)}t_{m-(m_0+1)}
\end{displaymath}
In terms of $n$
and for an arbitrary subset we have
\begin{equation}\label{sols}
s_{n}=s_{n_0}t_{\frac{n-q}{4}-m_0}-s_{n_0-4}t_{\frac{n-q}{4}-(m_0+1)}\,.
\end{equation}
We also conclude that if $a(n)$ is such a function that for all
$\epsilon$ there is an $m_0$ for which
$2-\epsilon <a(n)<2$ for all $n\geq m_0$
such that for large $n$ we have
\begin{equation}\label{cond}
a(n+4)\simeq a(n) \simeq a(n-4)
\end{equation}
and (\ref{sols}) with $a \to a(n)$ is an approximate solution to
$s_{n+4}-a(n)s_n+s_{n-4}=0$.

It should be noted that the bigger the value of $n$ is, the better
condition (\ref{cond}) is fulfilled. However, it does not imply that
$\omega(n)$, obtained after substituting $a$ by $a(n)$, is
constant. Rather, it varies very slowly compared to the change of
$s_n$ when $n$ is changed by four. One also must remember that
$a=a_0=2$ is a critical point of the asymptotic equation, so
substituting $a$ by two instead of observing condition (\ref{cond})
will not lead to correct results.

\subsection{Continued fractions}
\label{A:cont}

We will employ another tool used to investigate loop quantum cosmology
equations. This is a continued fraction, which we define by:
\begin{displaymath}
\frac{1}{a-\displaystyle\frac{\strut 1}{a-\cdots}}=x
\end{displaymath}
Noticing $(a-x)^{-1}=x$
we have
\begin{equation}\label{cfeq}
x^2-a x+1=0\,.
\end{equation}
This is the denominator of $G(x)$ in (\ref{g}) and so existence of the
continued fraction is related to the problem considered before. For
$|a|< 2$, Eq.~(\ref{cfeq}) has no solutions and so the continued
fraction does not converge. To see from the previous context why this
is the case we introduce $h(n+4)=s_{n+4}/s_n$ in (\ref{diffconst})
such that $h(n+4)=a-h(n)^{-1}$.
Thus,
\begin{displaymath}
\lim_{n\to\infty}h(n+4)=a-\frac{1}{a-\displaystyle\frac{\strut 1}{a-\cdots}}
\end{displaymath}
and one sees the close relationship between equation (\ref{diffconst})
and continued fractions. For $|a|< 2$ solutions are oscillating, and
so is $h(n+4)$, so the continued fraction does not exist. For $|a|
\geq 2$, on the other hand, $h(n+4)=a-x$
and the continued fraction is well-defined.

In particular, if we assume $a\geq 2$ then
$x=x_2$ from (\ref{roots}).
For finite $n$ and if sufficiently many steps are taken,
\begin{displaymath}
\frac{1}{a-\displaystyle\frac{\strut1}{a-
\displaystyle\frac{\strut1}{\cdots-h_0}}}
\end{displaymath}
is almost $h_0$ independent, and so very close to $x$.

\subsection{The case $|a|=2$}
\label{A:atwo}

We start with the solution to (\ref{diffconst}) for $a=2$,
$s_{n+4}-2s_n+s_{n-4}=0$ which is
\begin{displaymath}
s_{n+4}=(m+2)s_{n-4m}-(m+1)s_{n-4(m+1)}\,.
\end{displaymath}
We pick $m$ in a such way that we have $s_{n+4}$ in terms of initial
values at $n-4m_0$ and $n-4(m_0+1)$, say:
$s_{n+4}=(m_0+2)s_{n-4m_0}-(m+1)s_{n-4(m_0+1)}$.

To obtain $s_{n+8}$ in terms of the initial values, we shift the label
from $n$ to $n+4$ and choose $m$ such that $n+4-4m=n-4m_0$, i.e.\
$m=m_0+1$. Thus,
$s_{n+8}=(m_0+3)s_{n-4m_0}-(m_0+2)s_{n-4(m_0+1)}$
implying
\begin{displaymath}
s_{n+8}-s_{n+4}=s_{n-4m_0}-s_{n-4(m_0+1)}\,.
\end{displaymath}
Because on the right hand side we have initial values, the progress of
the sequence is constant. In the special case
$s_{n-4m_0}=s_{n-4(m_0+1)}$, the progress is zero and solutions are
constant. So solutions in the case $a=a_0=2$ are in general linearly
growing (except for one choice of initial values). This should be
compared with the oscillating nature for $|a|<2$ and exponential
growth for $|a|>2$. As we have already remarked, $a=a_0$ is a critical
point of the asymptotic equation, resembling critical points of
differential equations.

It is also interesting to discuss how the general procedure to find
bounded solutions
works here. For the subset $N_1$ we have
$s_{n+4}=(n/4+2)s_0-(n/4+1)s_{-4}$
from which the condition
\begin{displaymath}
s_0>\frac{n/4+1}{n/4+2}s_{-4}
\end{displaymath}
for a positive solution follows, and thus we must have $s_0>s_{-4}$ in
the limit $n \to \infty$. Using our previous approach with the
function $\rho$ and noting that
\begin{displaymath}
\frac{1}{2-\displaystyle\frac{\strut 1}{2-\cdots}}=1
\end{displaymath}
we have $r(m_0,q)=\lim_{n\to \infty} \rho_{a=2}(\bar{k}(n),n)^{-1}=1$,
reproducing $s_0>s_{-4}$.

The case $a=-2$ follows immediately after the substitution
$s_n=(-1)^{(n-q)/4} \bar{s}_n$.

\end{appendix}


\end{document}